\documentclass[aps,prl,reprint,groupedaddress,showpacs,nofootinbib]{revtex4-1}
\usepackage[utf8]{inputenc}
\usepackage{amsmath}
\usepackage{amssymb}
\usepackage{graphicx}
\usepackage[unicode=true,pdfusetitle,
 bookmarks=true,bookmarksnumbered=false,bookmarksopen=false,
 breaklinks=false,pdfborder={0 0 0},pdfborderstyle={},backref=false,colorlinks=false]
 {hyperref}

\renewcommand\[{\begin{equation}}
\renewcommand\]{\end{equation}}

\begin{document}
\global\long\def\d{\mathrm{d}}

\global\long\def\manifold{\mathcal{M}}

\global\long\def\lie{\mathcal{L}}

\title{Hamilton-Jacobi Formulation of the Thermodynamics of Einstein-Born-Infeld-AdS
Black Holes}

\author{Tekin Dereli}
\email{tdereli@ku.edu.tr}

\author{Kıvanç İ. Ünlütürk}
\email{kunluturk17@ku.edu.tr}

\affiliation{Department of Physics, Koç University, 34450 Sarıyer, İstanbul, Turkey}

\date{\today}

\begin{abstract}
A Hamilton-Jacobi formalism for thermodynamics was formulated by Rajeev
{[}Ann. Phys. 323, 2265 (2008){]} based on the contact structure
of the odd dimensional thermodynamic phase space. This allows one
to derive the equations of state of a family of substances by solving
a Hamilton-Jacobi equation (HJE). In the same work it was applied
to chargeless non-rotating black holes, and the use of Born-Infeld
electromagnetism was proposed to apply it to charged black holes as
well. This paper fulfills this suggestion by deriving the HJE for
charged non-rotating black holes using Born-Infeld theory and a negative
cosmological constant. The most general solution of this HJE is found.
It is shown that there exists solutions which are distinct from the
equations of state of the Einstein-Born-Infeld-AdS black hole. The
meaning of these solutions is discussed. 
\end{abstract}

\pacs{05.70.Ce, 02.40.Hw, 04.70.Bw}
\maketitle

\section{Introduction}

General relativity predicts that black holes obey the laws of thermodynamics
with the horizon area and the surface gravity playing the roles of
entropy and temperature, respectively \cite{Bardeen1973,Bekenstein1973}.
We may expect that in a quantum theory of gravitation these thermodynamic
quantities such as mass and area are treated as operators. It is therefore
reasonable to attempt to ``quantize'' thermodynamics. Another possible
application of quantized thermodynamics is to systems where the thermal
fluctuations are small, but the quantum fluctuations are not.

Thermodynamics can be formulated in terms of contact geometry \cite{Arnold1990},
where the first law is used to define a contact structure and the
equations of state pick out a Legendrian submanifold of the thermodynamic
phase space. With the aim of quantizing thermodynamics, a quantization
procedure for contact manifolds is established in \cite{Rajeev2008a}.
Subsequently, a Hamilton-Jacobi formalism for thermodynamics is developed
in \cite{Rajeev2008b}.

In this formalism one extends a thermodynamic system of $n$ degrees
of freedom into an $n$ parameter family; e.g. the ideal gas of fixed
particle number can be extended into the van der Waals family with
the parameters $a$ and $b$. This family can be described as a hypersurface
in the phase space, defined by the vanishing of a function which we
take to be the Hamiltonian. Given the Hamiltonian function $F$, one
can formulate a Hamilton-Jacobi equation (HJE), the characteristic
curves of which correspond to the dynamics generated by $F$. The
equations of state can be obtained in principle by solving the HJE.

In the same work the Hamilton-Jacobi formalism is applied to black
holes of one thermodynamical degree of freedom, i.e. to electrically
neutral non-rotating black holes. A negative cosmological constant
is introduced to extend the Schwarzschild black hole into the Schwarzschild-AdS
family. It is also proposed that the Born-Infeld action be used as
a modification of the Einstein-Maxwell equations to describe the family
of charged black holes (further modifications can be made to include
the rotating ones as well). Following this suggestion we shall apply
the Hamilton-Jacobi formalism to charged non-rotating black holes.

We first summarize Hamiltonian dynamics in contact geometry and how
it is applied to thermodynamics in \cite{Rajeev2008b}. Next we review
the non-rotating black hole solutions in the Einstein-Born-Infeld
(EBI) theory. We extract the thermodynamical quantities such as the
mass and the surface gravity and find the hypersurface in the thermodynamical
phase space which describes the EBIAdS family. Lastly, we write down
the HJE and discuss its solutions.

Throughout we work with units for which $c=G=4\pi\epsilon_{0}=1$.

\section{Hamilton-Jacobi Formalism for Thermodynamics}

Before describing the formalism that we shall use, we briefly review
contact geometry and contact Hamiltonian dynamics. The proofs of the
claims stated here and further reading on the subject can be found
in \cite{Geiges2009}.

\subsection{Contact Hamiltonian Dynamics}

A contact structure $\xi$ on a $2n+1$ dimensional manifold $\manifold$
is a codimension one distribution which is maximally non-integrable.
In other words it is (locally) the kernel of a 1-form $\alpha$ such
that
\begin{equation}
\alpha\wedge\left(\d\alpha\right)^{n}\neq0.\label{eq: contact condition}
\end{equation}
The contact form $\alpha$ is not unique since $\ker(f\alpha)=\ker(\alpha)$
if $f$ is a non-vanishing function. The contact condition (\ref{eq: contact condition})
is, however, independent of the choice of $\alpha$. For a fixed contact
form $\alpha$ there is a unique vector field $R_{\alpha}$ called
the \emph{Reeb vector field} which satisfies $i_{R_{\alpha}}\alpha=1$
and $i_{R_{\alpha}}\d\alpha=0$.

If a vector field $X$ generates a \emph{contactomorphism }$\manifold\rightarrow\manifold$,
i.e. a diffeomorphism which preserves the contact structure;
\[
\lie_{X}\alpha=\mu\alpha
\]
where $\mu$ is an arbitrary function, then $X$ is called a \emph{contact
vector field}. If the contact form $\alpha$ is fixed, there is a
one-to-one correspondence between the smooth functions $F:\manifold\rightarrow\mathbb{R}$
and the contact vector fields $X$ on $\manifold$ which is given
by
\begin{itemize}
\item $X\rightarrow F_{X}:=i_{X}\alpha,$
\item $F\rightarrow X_{F}$ defined as the unique solution of the equations
$i_{X_{F}}\alpha=F$ and $i_{X_{F}}\d\alpha=R_{\alpha}(F)\alpha-\d F$.
\end{itemize}
In this context the function $F$ is called the generating function
or the Hamiltonian and $X_{F}$ the corresponding Hamiltonian vector
field. Thus, in close analogy to symplectic geometry, a Hamiltonian
function $F$ defines a dynamics on $\manifold$ by the flow of $X_{F}$.
Note however that
\[
\lie_{X_{F}}F=FR_{\alpha}(F),
\]
so the Hamiltonian is in general not conserved. But if the initial
value of $F$  is zero, it remains zero.

As a final remark we note that if $L$ is a submanifold such that
$TL\subset\xi$, then $\dim L\leq n$. Such a submanifold of maximal
dimension $n$ is called a \emph{Legendrian submanifold}.

\subsection{Application to Thermodynamics}

The phase space $\manifold$ of a thermodynamic system of $n$ degrees
of freedom is $2n+1$ dimensional. For concreteness, take a gas with
a fixed number of particles. The variables $U,S,V,T$ and $p$ can
be taken as the coordinates on $\manifold$. The first law defines
a contact form
\[
\alpha=-\d U+T\d S-p\d V,
\]
whose kernel is the set of directions in which the energy is conserved.
The equations of state define an $n$ dimensional submanifold, all
of whose tangent vectors are annihilated by $\alpha$, i.e. a Legendrian
submanifold:
\[
U=U(S,V),\quad T=\left(\frac{\partial U}{\partial S}\right)_{V},\quad p=-\left(\frac{\partial U}{\partial V}\right)_{S}.
\]
Note that once the fundamental relation $U=U(S,V)$ is given, the
remaining $n$ equations of state follow automatically by the vanishing
of $\alpha$.

In general, let
\[
\alpha=du+p_{i}\d q^{i}
\]
 be a contact form on the thermodynamic phase space $\manifold$ with
coordinates $u$, $q^{1}$, $\dots$, $q^{n}$, $p_{1}$, $\dots$,
$p_{n}$\footnote{By Darboux's theorem one can always find such coordinates for a given
contact form $\alpha$.}. A family of substances can be described by allowing the fundamental
relation $u=\Phi(q^{1},\dots,q^{n})$ to depend on $n$ parameters\footnote{We \emph{choose} the number of parameters to be $n$ since this is
going to allow us to define a Hamiltonian function, see \cite{Rajeev2008b}.} $a_{1},\dots,a_{n}$ as well:
\begin{equation}
u=\Phi(q^{1},\dots,q^{n};a_{1},\dots,a_{n}).\label{eq: fundamental with parameters}
\end{equation}
E.g. for the van der Waals family we have $U=U(S,V;a,b)$, where $a$
and $b$ are the van der Waals parameters.

Given the fundamental relation (\ref{eq: fundamental with parameters})
and the other $n$ equations of state
\[
p_{i}=-\frac{\partial\Phi}{\partial q^{i}},
\]
we can eliminate the parameters $a_{1},\dots,a_{n}$ to get a single
relation between the $2n+1$ coordinates:
\[
F(u,q^{1},\dots,q^{n},p_{1},\dots,p_{n})=0.
\]
Given such a Hamiltonian $F$ for a family of substances, the equations
of state may be recovered by solving the first order PDE
\[
F(\Phi,q^{1},\dots,q^{n},-\frac{\partial\Phi}{\partial q^{1}},\dots,-\frac{\partial\Phi}{\partial q^{n}})=0.
\]
A complete integral of this PDE will depend on $n$ parameters $a_{1},\dots,a_{n}$.
The characteristic curves of this PDE are precisely the integral curves
of the Hamiltonian vector field $X_{F}$ \cite{Rajeev2008b,Courant1989}.

\section{The Einstein-Born-Infeld-AdS Black Hole}

The EBI field equations with a cosmological constant $\Lambda$ may
be derived from the action
\begin{equation}
S[e^{a},A]=\frac{1}{16\pi}\int\left(R^{ab}\wedge\star e_{ab}-2\Lambda\star1\right)+\int\mathcal{L}\star1,\label{eq: action}
\end{equation}
with the Born-Infeld Lagrangian \cite{Born1934}
\[
\mathcal{L}=\frac{1}{4\pi\lambda^{2}}\left(1-\sqrt{1-\lambda^{2}X-\lambda^{4}Y^{2}}\right),
\]
 where the quadratic invariants of the electromagnetic field are given
by
\[
X=\star\left(F\wedge\star F\right),\quad Y=\frac{1}{2}\star\left(F\wedge F\right).
\]
Here $R^{ab}$ are the curvature 2-forms of the Levi-Civita connection,
$e^{a}$ are the orthonormal coframe 1-forms, $e_{a}=\eta_{ab}e^{b}$
and $e_{a_{1}\cdots a_{r}}:=e_{a_{1}}\wedge\cdots\wedge e_{a_{r}}$.
The Born-Infeld parameter $\lambda$ is to be seen as a new fundamental
constant of nature. Note that in the limit $\lambda\to0$ the Born-Infeld
Lagrangian $\lie$ becomes the Maxwell Lagrangian $\lie\to(8\pi)^{-1}\star(F\wedge\star F)$.

The variational field equations for the case with no cosmological
constant are derived using the invariant tensor notation in \cite{Dereli2010a}.
The derivation with $\Lambda$ is analogous and here we simply note
the results. Variation with respect to the coframe $e$ leads to the
EBI field equations
\begin{equation}
-\frac{1}{2}R^{bc}\wedge\star e_{abc}+\Lambda\star e_{a}=8\pi\tau_{a},\label{eq: einstein-born-infeld eqn}
\end{equation}
with the stress-energy 3-forms
\[
\tau_{a}=M\star e_{a}+N\tau_{a}^{\text{(LED)}},
\]
where $M=\mathcal{L}-X\partial_{X}\mathcal{L}-Y\partial_{Y}\mathcal{L}$,
$N=8\pi\partial_{X}\mathcal{L}$ and $\tau_{a}^{\text{(LED)}}$ are
the stress-energy 3-forms of linear (i.e. Maxwell) electrodynamics:
\[
\tau_{a}^{\text{(LED)}}=\frac{1}{8\pi}\left(i_{a}F\wedge\star F-F\wedge i_{a}\star F\right).
\]
Variation with respect to the electromagnetic potential $A$ yields
the field equation $\d G=0$ where
\[
G=\frac{1}{4\pi\sqrt{\Delta}}\left(\star F+\lambda^{2}YF\right).
\]

The static spherically symmetric solution without the cosmological
constant was first found by Hoffmann \cite{Hoffmann1935} and then
rediscovered by Demianski \cite{Demianski1986}. The solution in the
presence of a cosmological constant was noted  in \cite{Fernando2003}.
Note that the action they use differs from ours by the absence of
the $Y^{2}$ term. However we shall see below that in the static case
one has $Y=0$, hence our results agree. The solution is given by
\begin{align*}
g & =-f\left(r\right)\d t^{2}+\frac{\d r^{2}}{f\left(r\right)}+r^{2}\left(\d\theta^{2}+\sin^{2}\theta\d\phi^{2}\right),\\
F & =\frac{Q}{\sqrt{r^{4}+a^{4}}}\d r\wedge\d t,
\end{align*}
where
\begin{align}
f\left(r\right) & =-\frac{\Lambda}{3}r^{2}+1-\frac{2M}{r}+\frac{2Q^{2}}{ar}h\left(\frac{r}{a}\right),\label{eq: f(r)}\\
h\left(x\right) & :=\int_{x}^{\infty}\d y\left(\sqrt{y^{4}+1}-y^{2}\right),
\end{align}
$Q$ is the total electric charge, $a=(\lambda|Q|)^{1/2}$ and $M$
is the total mass\footnote{For the definition of the total mass in an asymptotically (A)dS spacetime,
see \cite{Abbott1982}}. 
\begin{figure}
\includegraphics[width=8cm]{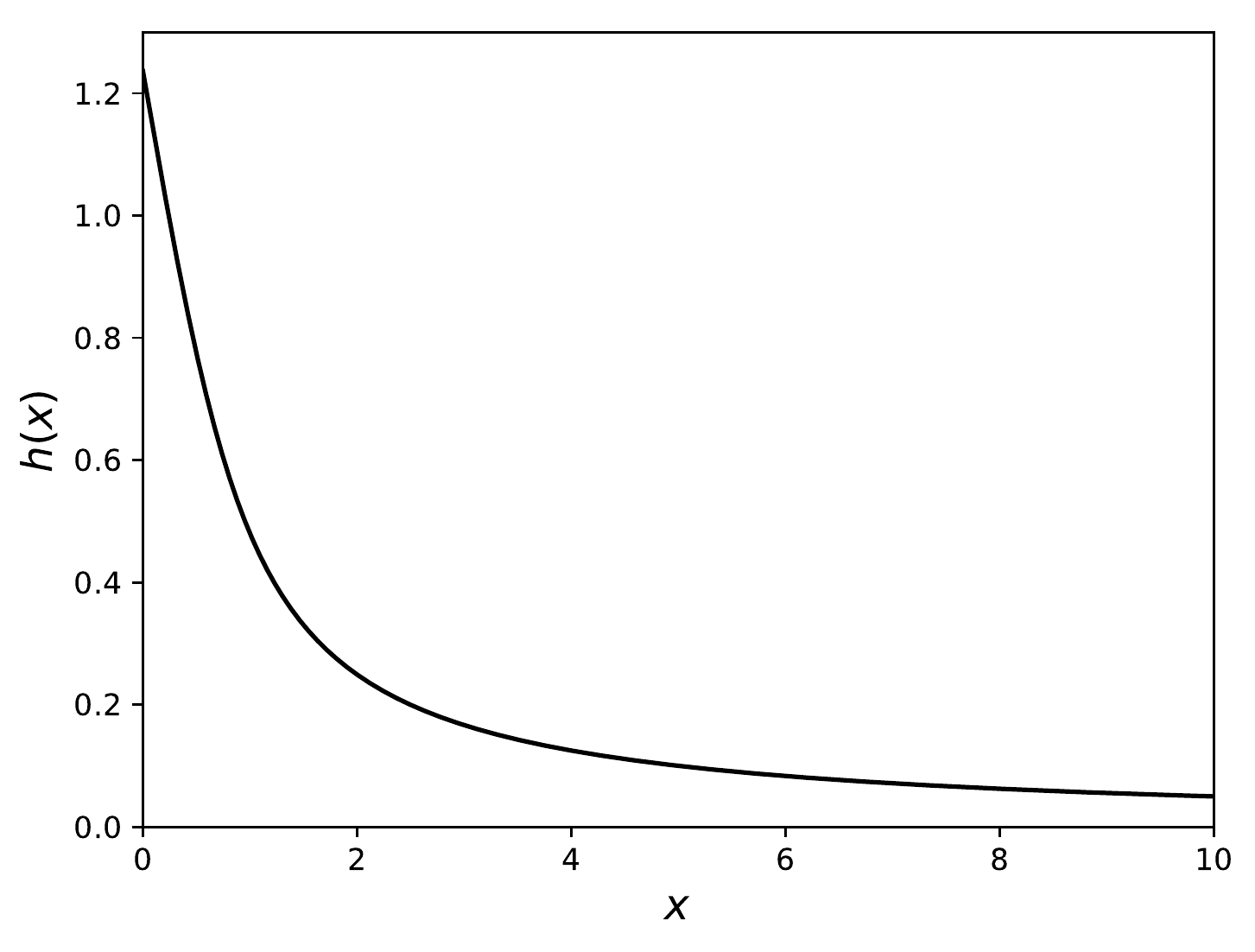}
\caption{Plot of the function $h(x)$ defined in (\ref{eq: f(r)}). It satisfies
$h(0)\approx1.24$ and $\lim_{x\to\infty}h(x)=0$.}

\end{figure}
For simplicity we shall work with a negative cosmological constant
$\Lambda=-3l^{-2}$ as in \cite{Rajeev2008b}. Then $f$ can be expanded
for $a\ll r$ as
\[
f(r)=\frac{r^{2}}{l^{2}}+1-\frac{2M}{r}+\frac{Q^{2}}{r^{2}}\left[1-\frac{a^{4}}{20r^{4}}+\mathcal{O}\left(\frac{a^{8}}{r^{8}}\right)\right].
\]
In particular the spacetime is asymptotically AdS and in the limit
$\lambda\to0$ we recover the Reissner-Nordström-AdS (RNAdS) black
hole as expected.

\begin{figure}
\includegraphics[width=8cm]{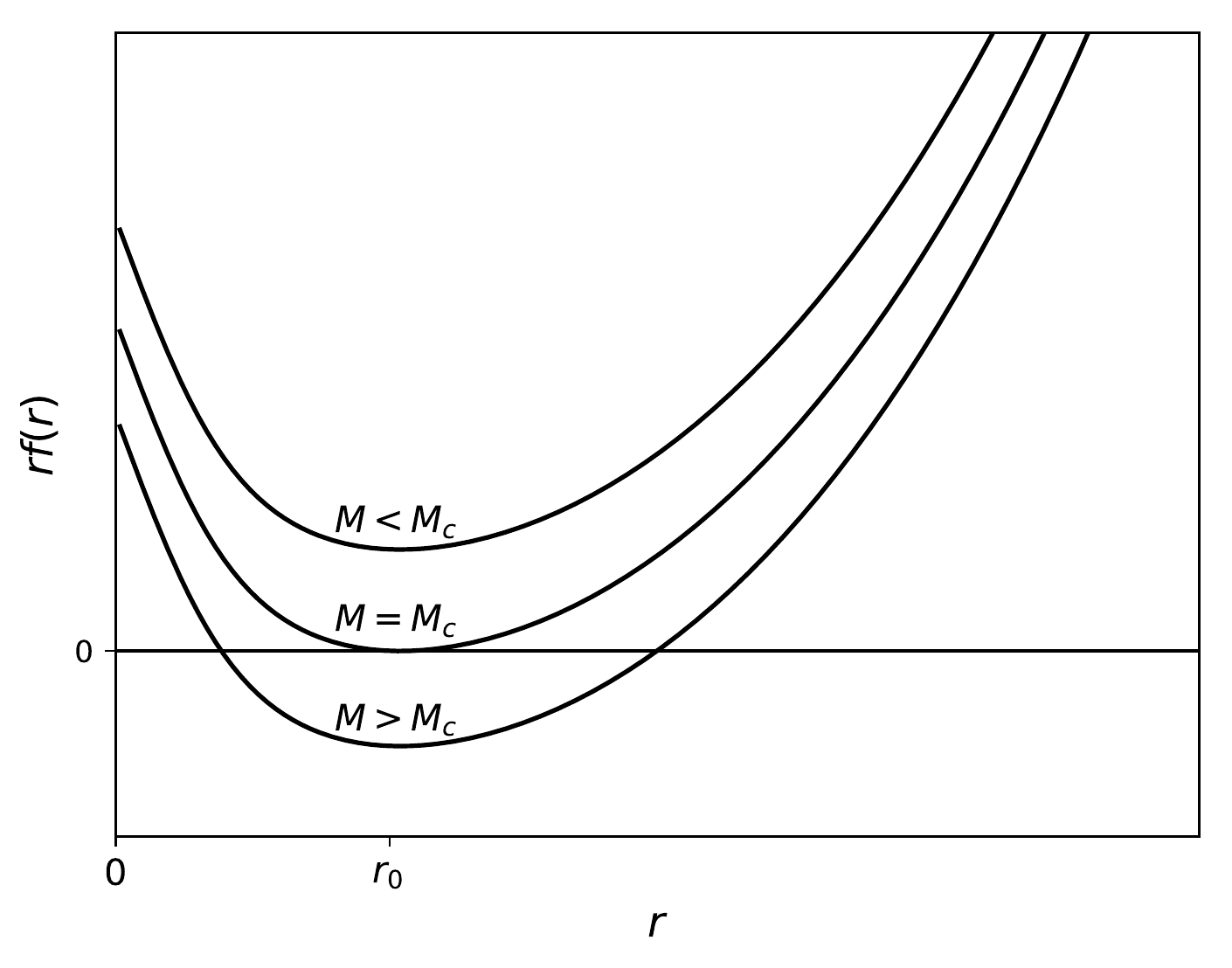}
\caption{Typical plots of $rf(r)$ for the three cases $M<M_{\text{c}}$, $M=M_{\text{c}}$
and $M>M_{\text{c}}$. The number of horizons is different in each
case, analogous to the RN black hole. The minimum occurs at $r_0$,
which is independent of $M$.}
\label{f plot}
\end{figure}

To investigate the horizon structure we first note that
\[
\lim_{r\rightarrow\infty}rf(r)=\infty\quad\text{and}\quad\left.rf(r)\right|_{r=0}=-2M+\frac{2Q^{2}}{a}h(0)
\]
($h\left(0\right)\approx1.24$ is finite.). We shall limit our attention
to the case
\begin{equation}
2M<\frac{2Q^{2}}{a}h(0)\quad\text{and}\quad\lambda<2Q,\label{eq: assumptions}
\end{equation}
which includes the RNAdS black hole. With the assumptions in (\ref{eq: assumptions}),
the function $rf(r)$ has exactly one minimum whose position we denote
by $r_{0}=r_{0}(\lambda,\Lambda,Q)$. Since $rf(r)$ is positive at
$r=0$ and $r=\infty$, it will have no zeros, a double zore or two
zeros if $r_{0}f(r_{0})$ is positive, zero or negative, respectively.
Hence from (\ref{eq: f(r)}) we see that there is a critical mass
\[
2M_{\text{c}}=\frac{r_{0}^{3}}{l^{2}}+r_{0}+\frac{2Q^{2}}{a}h(r_{0}/a),
\]
such that (see Fig. \ref{f plot}) if
\begin{enumerate}
\item $M<M_{\text{c}}$, there is no horizon. In that case there is a naked
singularity at $r=0$ where the Kretschmann scalar $\mathcal{K}=2\star(R_{\phantom{a}b}^{a}\wedge\star R_{\phantom{b}a}^{b})$
diverges.
\item $M=M_{\text{c}}$, we have an extremal black hole with one horizon.
\item $M>M_{\text{c}}$, there are two horizons.
\end{enumerate}
Thus the situation is completely analogous to the RN case.

If $l$ is small, it may not be possible to satisfy the condition
$M>M_{\text{c}}$ simultaneously with (\ref{eq: assumptions}), but
it is possible if $l$ is sufficiently large. In the subsequent discussion
we shall assume that $M>M_{\text{c}}$, which should cover the physically
relevant cases. We denote the outer horizon radius by $r_{\text{H}}$.

The thermodynamical equations of state were given in \cite{Rasheed1997}
for the case with $\Lambda=0$ and then in \cite{Dey2004} for a non-zero
$\Lambda$. In the following we note these equations and derive from
them the HJE.

Using (\ref{eq: f(r)}) and the fact that $f(r_{\text{H}})=0$ we
can write the mass of the black hole as
\begin{equation}
2M=\frac{r_{\text{H}}^{3}}{l^{2}}+r_{\text{H}}+\frac{2Q^{2}}{a}h(r_{\text{H}}/a).\label{eq: m}
\end{equation}
The surface gravity $\kappa=(1/2)\left.\d f/\d r\right|_{r=r_{\text{H}}}$
is given by
\begin{align}
\kappa & =\frac{r_{\text{H}}}{l^{2}}+\frac{M}{r_{\text{H}}^{2}}-\frac{Q^{2}}{ar_{\text{H}}^{2}}h(r_{\text{H}}/a)-\frac{Q^{2}}{a^{4}r_{\text{H}}}\left(\sqrt{r_{\text{H}}^{4}+a^{4}}-r_{\text{H}}^{2}\right),\label{eq: kappa}
\end{align}
and the electrostatic potential on the horizon is
\begin{equation}
\Phi=\int_{r_{\text{H}}}^{\infty}\frac{Q}{\sqrt{x^{4}+a^{4}}}dx.\label{eq: phi}
\end{equation}
Introducing the surface area $A=4\pi r_{\text{H}}^{2}$ and using
the identity 
\[
h(x)=\frac{2}{3}\int_{x}^{\infty}\frac{\d y}{\sqrt{y^{4}+1}}-\frac{x}{3}\left(\sqrt{x^{4}+1}-x^{2}\right),
\]
 it is straightforward to verify the first law 
\[
\d M=\frac{\kappa}{8\pi}\d A+\Phi\d Q,
\]
 which defines our contact structure. Furthermore, from the three
equations (\ref{eq: m})-(\ref{eq: phi}) we can eliminate $a$ and
$l$ to get a single relation between the variables $M$, $A$, $\kappa$,
$Q$ and $\Phi$:
\[
3M-\frac{\kappa A}{4\pi}-2\Phi Q=\sqrt{\frac{A}{4\pi}}.
\]
We therefore see –using the first law– that the EBIAdS family is described
by the hypersurface $F(M,A,Q,p_{A},p_{Q})=0$ with the Hamiltonian
\[
F(M,A,Q,p_{A},p_{Q})=3M-2p_{A}A-2p_{Q}Q-\sqrt{\frac{A}{4\pi}}.
\]
The HJE we get from this Hamiltonian is
\begin{equation}
3M-2A\frac{\partial M}{\partial A}-2Q\frac{\partial M}{\partial Q}=\sqrt{\frac{A}{4\pi}}.\label{eq: hje}
\end{equation}
This is a first order quasi-linear PDE which is surprisingly nice
and we can do even better than finding a particular complete integral.
One can indeed show that the most general solution is
\[
2M=\sqrt{\frac{A}{4\pi}}+\left(\frac{A}{4\pi}\right)^{3/2}u(4\pi Q/A),
\]
where $u$ is an arbitrary function which must be fixed by a boundary
condition. The complete integral corresponding to the actual equation
of state (\ref{eq: m}) is given by the choice
\[
u\left(x\right)=\frac{1}{l^{2}}+\frac{2x^{3/2}}{\lambda^{1/2}}h\left(\frac{1}{\sqrt{\lambda x}}\right).
\]

It should be noted, however, that (\ref{eq: m}) is not the only complete
integral of the PDE (\ref{eq: hje}) as, e.g., the choice $u(x)=l^{-2}+2x^{3/2}\lambda^{-1/2}$
also yields a complete integral. It is not exactly clear what this
non-uniqueness means, but at the very least it shows that one must
be careful to use the HJE to get the equations of state of a family
of substances.

\section{Conclusion}

To study the thermodynamic HJE for charged black holes we have made
the RN black hole into a two parameter family by introducing a (negative)
cosmological constant and replacing the Maxwell Lagrangian by the
Born-Infeld one. Under the assumption that the Born-Infeld parameter
$\lambda$ is sufficiently small, the horizon structure of the resulting
static black hole is quite similar to that of the RN one. By this
we mean that there is a critical mass below which there is no horizon
(Fig. \ref{f plot}).

An element of the EBIAdS family has two thermodynamical degrees of
freedom. Hence its phase space is five dimensional, which can be coordinatized
by $M$, $\kappa$, $A$, $\Phi$ and $Q$, and the two parameters
$\Lambda$ and $\lambda$ specify the particular element. Using the
three equations of state (\ref{eq: m})-(\ref{eq: phi}) we were able
to eliminate the parameters $\Lambda$ and $\lambda$ to get a single
relation between the phase space variables. This relation defines
a hypersurface by the vanishing of a Hamiltonian function and therefore
yields a HJE.

The HJE (\ref{eq: hje}) we get for the EBIAdS family is quasi-linear
and we can find an analytic expression for the most general solution.
It is interesting to note that the solution is far more general than
the equation of state of the EBIAdS black hole. A function must be
specified by a boundary condition to get the actual equation of state,
not just two constants of integration. As we mentioned, the precise
meaning of the existence of these solutions that do not correspond
to the actual equation of state is not clear. This may get clearer
if it is understood whether and how the HJE is related to the quantization
of black holes. In any case it provides a caveat against the use of
the Hamilton-Jacobi formalism to determine the equations of state
of a family of substances.

It should also be remarked that the above is not the only way of extending
the RN black hole into a two parameter family; one may think of parameters
other than the cosmological constant and the Born-Infeld parameter.
In fact, we can introduce $n$ parameters to a system of $n$ degrees
of freedom in a completely arbitrary manner. It is possible that there
is another choice of extension which is free of this ambiguity. Moreover,
the question of finding a HJE for rotating black holes is still open.

\begin{acknowledgments}
We thank C. Yetişmişoğlu for valuable discussions.
\end{acknowledgments}

\bibliographystyle{apsrev}

\end{document}